\documentclass[prl,twocolumn,twoside,showpacs,floatfix]{revtex4}
\usepackage{graphics,amssymb,amsmath}

\begin{document}
\title{Enhancement of Coherent Response by Quenched Disorder}

\author{H. Hong}
\affiliation{Department of Physics, Chonbuk National University, Jeonju 
561-756, Korea}

\date{\today}

\begin{abstract}
We investigate the effects of quenched disorder on the coherent response in a driven system of 
coupled oscillators.  In particular, the interplay between quenched noise and periodic 
driving is explored, with particular attention to the possibility of resonance. 
The phase velocity is examined as the response of the system; revealed is the enhancement of 
the fraction of oscillators locked to the periodic driving, displaying resonance behavior.  
It is thus concluded that resonance behavior may also be induced by quenched 
disorder which does not have time-dependent correlations. 
 
\end{abstract}
\maketitle
In recent years, the {\it stochastic resonance} (SR) has drawn much 
attention, which stands for the phenomena that the response of a system to 
a periodic driving force is enhanced by an appropriate amount of noise rather than 
suppressed~\cite{ref:SR1,ref:SR2,ref:SR3}.
Those SR phenomena observed in various real 
systems~\cite{ref:Schmitt,ref:ringlaser,ref:crayfish} are 
known to occur through the cooperative interplay
between the noise and the external driving force, and require 
three ingredients: (i) an energetic activation barrier; (ii) a weak input such 
as periodic driving; (iii) a random noise~\cite{ref:SR3}.
In particular, the SR may be understood in terms of matching of two 
time scales, the period of the driving force and the 
inverse of the Kramers hopping rate associated with the random 
noise.  Note that the noise here usually has time-dependent correlations.
Namely, the correlation between two noise forces 
$\eta_i$ and $\eta_j$ is given by 
$\langle \eta_i(t)\eta_j (t')\rangle= 2T\delta_{ij}\delta(t-t')$, where 
$T$ represents the noise strength.
Here naturally arises a question how does the SR behavior 
depends on such characteristics of the noise. 
In particular, one may ask whether such resonance behavior 
appears even for the quenched noise without time-dependent correlations.   

To resolve this, we consider the system of coupled 
oscillators described by 
\begin{equation}
\dot{\varphi_i}= \omega_i - \frac{K}{N}\sum_{j=1}^{N}\sin(\varphi_i-\varphi_j) 
+ I_i \cos\Omega t,
\label{eq:model}
\end{equation}
where $\varphi_i$ represents the phase of the $i$th limit-cycle oscillator.  
The first term $\omega_i$ on the right-hand side denotes the 
intrinsic frequency of the $i$th oscillator, assumed to be 
distributed according to the Gaussian distribution 
$g(\omega)$ with zero mean $(\langle \omega \rangle=0)$ 
and correlations $\langle \omega_i \omega_j \rangle= D\delta_{ij}$.
Note that correlations between $\omega_i$'s are not time-dependent
but quenched in time, in sharp contrast with the usual thermal noise.
The second term on the right-hand side describes the coupling 
between oscillators, whereas the third term 
represents the external periodic driving with
the driving strength $I_i$ chosen from a 
certain distribution function $f(I)$.  
In the absence of the external driving $(I_i=0)$, Eq.~(\ref{eq:model}) 
describes the well-known Kuramoto model, where the synchronization 
phenomena have been investigated extensively~\cite{ref:Kuramoto}.
The scattered intrinsic frequency or quenched noise 
$\omega_i$ competes with the coupling strength $K$ in the system: When the 
coupling is strong enough to overcome the scatteredness of 
intrinsic frequencies, synchronization emerges.  Such 
synchronization behavior may be explored by measuring the 
order parameter 
\begin{equation}
\Delta\equiv \Biggl\langle \Biggl|\frac{1}{N}\sum_{j}e^{i\varphi_j}
\Biggr|\Biggr\rangle,
\label{eq:def_order}
\end{equation}
where nonzero $\Delta$ implies the emergence of synchronization.
It is known that the synchronization behavior of the system 
is characterized by $\Delta\sim (K-K_c)^{\beta}$ with $\beta=1/2$ 
near the critical coupling strength $K_c=2/\pi g(0)$~\cite{ref:Kuramoto}.  
The Kuramoto model has been also extended by means of introducing a constant
in the argument of sine coupling~\cite{ref:Kuramoto_Sakaguchi}.  Such 
extended model is also studied in Ref.~\cite{ref:Wiesenfeld_Colet_Strogatz}, 
where frequency locking in the system of Josephson arrays is investigated.

Meanwhile, when the external periodic driving comes into system, the 
synchronization behavior has been observed to appear 
periodically~\cite{ref:period_synch}. 
However, the possibility of the resonance phenomena, which 
may occur via the cooperative interplay between the external driving and 
the quenched noise, has not 
been investigated.  
In this work, we examine the linear response of the system, with particular 
attention to the possibility of the resonance behavior due to 
the quenched disorder. 

We first investigate analytically the dynamics of the 
system governed by Eq.~(\ref{eq:model}). 
The order parameter $\Delta$ defined in Eq.~(\ref{eq:def_order}) allows us 
to reduce Eq.~(\ref{eq:model}) into a single equation
\begin{equation}
\dot\varphi = \omega - K\Delta\sin \varphi + I\cos\Omega t,
\label{eq:single}
\end{equation} 
where $\Delta$ is to be determined by imposing self-consistency, and 
indices are suppressed for simplicity. 
Equation~(\ref{eq:single}) reminds one of the single resistively shunted 
Josephson junction under combined direct and alternating currents.
It is well known that such a system can be locked to the external 
driving, which is characterized by the Shapiro step~\cite{ref:Shapiro}
\begin{equation}
\frac{\bar{\dot\varphi}}{\Omega}= n 
\end{equation}
with $n$ integer, where $\bar{\dot{\varphi}}~(\equiv v)$ is the time-averaged phase velocity.
Such mode locking features suggest the ansatz 
\begin{equation}
\varphi=\varphi_0+n\Omega t + \sum_{p=1}^{\infty} A_{p}
\sin(p\Omega t + \alpha_{p})
\label{eq:ansatz}
\end{equation}
for the locked phase of the oscillator on the $n$th step.
The external periodic driving in the system 
also leads us to expect periodic synchronization~\cite{ref:period_synch}, where 
the order parameter $\Delta$
is decomposed as  
\begin{equation}
\Delta=\Delta_0 + \sum_{q=1}^{\infty}\Delta_{q} 
\cos(q\Omega t + \beta_{q}).
\label{eq:delta_ansatz}
\end{equation}
Inserting Eqs.~(\ref{eq:ansatz}) and (\ref{eq:delta_ansatz}) into 
Eq.~(\ref{eq:single}), we obtain
\begin{eqnarray}
n\Omega+\sum_{p=1}^{\infty}p\Omega A_{q} \cos(p\Omega t + \alpha_{p})
- I\cos\Omega t 
\nonumber\\
= \omega - K\Biggl[\Delta_0 + \sum_{q=1}^{\infty}\Delta_q\cos(q\Omega t+\beta_q)
\Biggr] \nonumber\\
\times\Biggl(\prod_{r=1}^{\infty}
\sum_{\ell_{r}=-\infty}^{\infty} J_{\ell_{r}}(A_{r})\Biggr)
\sin \Phi,
\end{eqnarray}
where $J_{\ell_{r}}(x)$ is the $\ell_{r}$th Bessel function and 
$\Phi\equiv \varphi_0 + \sum_{s=1}^{\infty}\ell_{s}\alpha_{s}
+(n+\sum_{s=1}^{\infty}s\ell_s)\Omega t$.  
The integers $\ell_{s}$ satisfying $\sum_{s}s\ell_s= -n$ contribute to the dc 
component in $\Phi$:
\begin{equation}
n\Omega=\omega-K\Delta_0\Biggl(\prod_{r=1}^{\infty}
{\sum_{\ell_r}}^{'}J_{\ell_{r}}(A_{r})\Biggr)
\sin \tilde{\Phi} +{\cal{O}}(K\Delta_1),
\label{eq:dc_com}
\end{equation}
where $\tilde{\Phi}\equiv \varphi_0 + \sum_{s=1}^{\infty}\ell_{s}\alpha_{s}$ and 
the prime in the summation stands for the constraint 
$\sum_{s}s\ell_s = -n$.  This gives an estimation of the dc driving 
strength $\omega$ corresponding to the integer locking.  The amplitude 
$A_p$ and phase $\alpha_p$ of the ac component with 
frequency $p\Omega$ can be determined from the equation
\begin{eqnarray}
&-&p\Omega A_p \cos(p\Omega t +\alpha_p)+\delta_{p,1}I\cos p\Omega t \nonumber\\
&=&K\Delta_0 \Biggl(\prod_{r=1}^{\infty}
\sum_{\ell^{+}_r} J_{\ell^{+}_{r}}(A_r)\Biggr)\sin\Biggl(
\varphi_0+p\Omega t + \sum_{s=1}^{\infty}
{\ell^{+}_s}\alpha_s \Biggr) \nonumber\\
&+&K\Delta_0\Biggl(\prod_{r=1}^{\infty}\sum_{\ell^{-}_r} J_{\ell^{-}_r}(A_r)
\Biggr) \sin\Biggl(\varphi_0-p\Omega t + \sum_{s=1}^{\infty}
{\ell^{-}_s}\alpha_s \Biggr) \nonumber\\
&+&K\Delta_p \cos(p\Omega t+\beta_p)
\Biggl(\prod_{r=1}^{\infty}{\sum_{\ell_r}}^{'}J_{\ell_{r}}(A_{r})\Biggr) 
\sin {\tilde{\Phi}}
\label{eq:ac_comp}
\end{eqnarray}
with integers $\ell^{+}_r$ and $\ell^{-}_r$ satisfying 
$\sum_{s=1}^{\infty} s \ell^{+}_s=p-n$
and $\sum_{s=1}^{\infty} s \ell^{-}_s=-p-n$, respectively.
When $K\Delta_0$ is sufficiently small compared with 
the driving amplitude, 
Eq.~(\ref{eq:ac_comp}) with $p=1$ yields 
$A_1$ and
$\alpha_1$ to the zeroth order in $K\Delta_0$: $A_1 = I/\Omega$ and 
$\alpha_1=0$.  
It is observed numerically that the dc component $\Delta_0$ of the 
order parameter is much larger than 
higher-order (ac) components $[\Delta_0 \gg \Delta_{\ell} (\ell \geq 1)]$, which 
allows the expansion of the locked phase to the zeroth order in $K\Delta_0$. 
This gives the (locked) phase of the 
oscillator on the $n$th step as 
\begin{equation}
\varphi= \varphi_0+n\Omega t + \frac{I}{\Omega}\sin\Omega t+
{\cal{O}}(K\Delta_0 I/2\Omega^2), 
\label{eq:lockedphase}
\end{equation}
which yields the range of locked oscillators: 
\begin{equation}
\omega - n\Omega = K\Delta_0 (-1)^n J_n (I/\Omega)
\sin\phi_0 + {\cal{O}}(K\Delta_1).
\end{equation} 
It implies that the oscillators in the range
\begin{equation}
n\Omega - K\Delta_0 |J_n(I/\Omega)|\leq \omega 
\leq n\Omega + K\Delta_0 |J_n(I/\Omega)|
\label{eq:lockingcondition}
\end{equation}
display the locking behavior 
$(v/\Omega=n)$, with the higher-order terms of
${\cal O}(K\Delta_1)$ neglected. 

The quenched disorder $\omega$ and the driving amplitude $I$ 
are chosen from the distribution function $g(\omega)$ and $f(I)$, respectively, 
which yields the fraction $r_n$ of the oscillators locked to the 
$n$th step: 
\begin{equation}
r_n = 
\int_{-\infty}^{\infty} f(I) dI 
\int_{n\Omega-K\Delta_0 |J_n(I/\Omega)|}^{n\Omega+K\Delta_0 |J_n(I/\Omega)|}
g(\omega) d\omega.
\label{eq:fn_1}
\end{equation}
For the Gaussian distribution 
$g(\omega)=(1/\sqrt{2\pi D})e^{-{\omega^2/2 D}}$ and the delta function 
one $f(I)=(1/2)[\delta(I-I_0)+\delta(I+I_0)]$, the fraction 
$r_n$ in Eq.~(\ref{eq:fn_1}) is given by
\begin{eqnarray}
r_n &=& \frac{1}{2} \Biggl[
{\rm{erf}}\Biggl(\frac{n\Omega+K\Delta_0 |J_n(I_0/\Omega )|}{\sqrt{2 D}}\Biggr) 
\nonumber\\
~~~~~~~~&-& {\rm erf}\Biggl(\frac{n\Omega-K\Delta_0 |J_n(I_0/\Omega )|}{\sqrt{2 D}}\Biggr)
\Biggr],
\label{eq:fn}
\end{eqnarray}
where ${\rm erf}(x)$ denotes the error function.
To estimate the behavior of the fraction $r_n$ 
as the variance $D$ varies, we should know the behavior of the 
dc component $\Delta_0$ as a function of the variance $D$.  
Note that the self-consistency 
equation for the order parameter gives the behavior of $\Delta_0$ only 
near the critical point.  
To obtain $\Delta_0$ in the whole range of the disorder strength, we 
resort to numerical simulations and integrate Eq.~(\ref{eq:model}) 
via Heun's method~\cite{ref:Heun} with the discrete time step $\delta t = 0.01$.
While the equations of motion are integrated for $N_t=8\times 10^4$ 
time steps, the data from the first $N_t/2$ steps are discarded in 
measuring quantities of interest.
The system size $N$ has been considered up to $N=20000$, so that no 
appreciable size-dependence is observed.
The driving amplitude $I_0$ in the distribution 
$f(I)$ and the driving frequency $\Omega$ have been chosen to be 
$I_0=0.8$ and $\Omega=1.0$, $1.2$, and $1.4$, respectively; the 
coupling strength $K$ has been set equal to unity for 
convenience.  
We measure the order parameter $\Delta$, and obtain the 
component $\Delta_0$ by taking the time average.

\begin{figure}
\centering{\resizebox*{!}{6.0cm}{\includegraphics{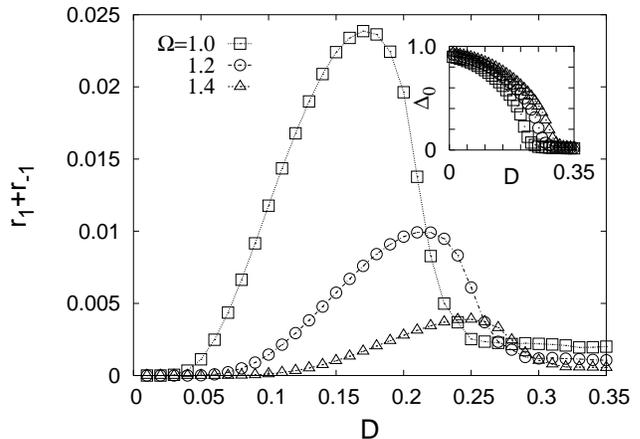}}}
\caption{Total fraction $r_1 + r_{-1}$ of the oscillators locked to the 
first step ($n=\pm 1$) versus the disorder strength 
$D$ for $K=1.0,~I_0=0.8$, and $\Omega=1.0,~1.2$, and $1.4$.
Inset: dc component $\Delta_0$ in the system of size $N=20000$ at various 
driving frequencies.
} 
\label{fig:f1}
\end{figure}

The dc component $\Delta_0$ is shown in the inset of 
Fig.~\ref{fig:f1} as a function of the disorder strength $D$, displaying 
the monotonically decreasing behavior. 
Using this, we compute the fraction $r_{\pm 1}$ of the 
oscillators which are locked to the first step since $r_{\pm 1}$ is 
most dominant over other components $[r_{\ell} (\ell \geq 2)]$.
Figure~\ref{fig:f1} displays the total fraction $r_{1}+r_{-1}$ for 
$K=1.0,~I_0=0.8$, and $\Omega=1.0, 1.2, 1.4$ versus the disorder strength $D$. 
It is found that the fraction first increases with $D$, which implies 
that a larger number of oscillators tends to follow the 
driving force as the quenched disorder becomes stronger. 
Remarkably as the disorder is increased further, the fraction reaches the 
maximum and begins to decrease.
For example, the optimal disorder strength is observed to be 
$D_m \approx 0.17$ for $\Omega=1.0$.
Such behavior of the fraction $r_{\pm 1}$ is reminiscent of the 
SR, which is known to occur through the cooperative interplay 
between the external periodic driving force and the random noise.
Note that in sharp contrast to the random noise in the 
conventional SR, the noise in the system governed 
by Eq.~(\ref{eq:model}) is quenched with no time-dependent correlations.
Namely, here the quenched disorder enhances the coherent response of the system.
It is also shown that as the driving frequency $\Omega$ is raised, the fraction $r_{\pm 1}$ 
of locked oscillators diminishes while the optimal strength $D_m$ shifts to larger
values.  These tendencies reflect that the oscillators are 
reluctant to follow the driving which changes too fast, and that 
the disorder strength $D$ should be enlarged to fit the high driving frequency. 
Another point is that the fraction $r_0$ of the oscillators 
locked to the $n=0$ step does not exhibit such resonance 
behavior; it rather displays the monotonic decreasing behavior, which is 
quite similar to that of the dc component $\Delta_0$ of 
the phase order parameter.  
Higher fractions for $n\geq 2$ also show the 
resonance behavior, although the magnitude is too 
small to be clearly discriminated.  
We have also investigated the role of coupling on the enhancement of coherent 
response.  As the coupling strength $K$ decreases, such resonance behavior is 
found to be suppressed.

\begin{figure}
\centering{\resizebox*{!}{6.0cm}{\includegraphics{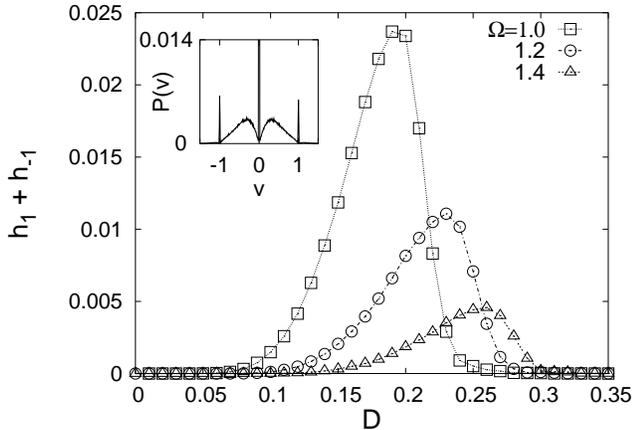}}}
\caption{Total height $h_1+h_{-1}$ of the peak 
at $v=\pm \Omega$ in the probability distribution versus
the disorder strength $D$.  Inset: probability distribution 
$P(v)$ of the time-averaged phase velocity $v$ 
for $K=1.0,~I_0=0.8,~\Omega=1.0$, and $D=0.20$.}
\label{fig:histo_reso}
\end{figure}

To confirm such resonance behavior, we now perform numerical simulations.  
We obtain the probability distribution $P(v)$ 
for the time-averaged phase velocity $v$ under the same 
conditions as before, varying 
the disorder strength $D$. 
The data are then averaged over 100 independent 
sets of $\{\omega_i\}$.
The inset of Fig.~\ref{fig:histo_reso} displays the probability 
distribution for $\Omega=1.0$ and $D=0.20$, where three sharp peaks appear at 
$v=0$ (not fully shown) and $\pm \Omega$.  
The two peaks at the driving frequency $(\pm\Omega)$ manifest that some 
fraction of the oscillators follow the external driving force, locked to the 
latter.  
Note that such peaks may also appear at higher 
frequencies $n\Omega~(|n| \geq 2)$, although they are too small to be 
manifestly shown. 

In the absence of the quenched disorder ($D=0$), those 
peaks at $v=\pm \Omega$ do not emerge, and only 
the delta peak appears at $v=0$. 
As the disorder strength is increased further, on the other hand, peaks 
show up at the driving frequency, with the height growing.  
The peak height $h_{\pm 1}$ at the driving frequency may be regarded as 
a good indicator which describes the response of the system to 
the external driving.

We numerically obtain the peak height, subtracting the background noise given by 
the average of ten data points around the peak.
Figure~\ref{fig:histo_reso} displays the total height 
$h_1+h_{-1}$ versus the disorder strength $D$ 
in the system of size $N=20000$, with the driving 
frequency varied from $\Omega=1.0$ to $\Omega=1.4$.  
We have considered the system size up to $N=40000$, where 
no appreciable size-dependence is observed.  
The total height is found to first increase with the disorder strength $D$
and reaches its maximum at a finite value of the disorder, displaying quite a similar 
feature to the fraction $r_{\pm 1}$ [see Fig.~\ref{fig:f1}].  
The enhancement of the height indicates that the quenched disorder 
actually increases the number of the oscillators locked to the 
external driving.
To our knowledge, such enhancement induced by the quenched disorder 
without time-dependent correlations has not been addressed before.  
It is observed in Fig.~\ref{fig:histo_reso} that the increase of the driving
frequency tends to suppress the total height, and to shift the optimal 
disorder strength to larger values, which may be related with the intrinsic
time scale of the system.  
Note that the conventional SR phenomena have been known to occur when the 
Kramers hopping rate and the external driving frequency match each other.  
To see such time scale matching in the quenched-noise-induced resonance, we 
investigate the relaxation dynamics of the system, and 
probe the time evolution of the renormalized synchronization order parameter
\begin{equation}
{\tilde{\Delta}(t)}\equiv \frac{\Delta(t)-\Delta_{{\rm{eq}}}}  
{\Delta(0)-\Delta_{{\rm{eq}}}}, 
\label{eq:renormDelta}
\end{equation}
where $\Delta_{\rm{eq}}$ and $\Delta(0)$ represent the equilibrium value 
and the initial one, respectively.  
The renormalized order parameter $\tilde{\Delta}(t)$ is thus expected  
to decay from unity ($t=0)$ to zero ($t\rightarrow \infty$): 
$\tilde{\Delta}(t) \sim {\rm exp}(-t/\tau)$. 

\begin{figure}
\centering{\resizebox*{!}{6.0cm}{\includegraphics{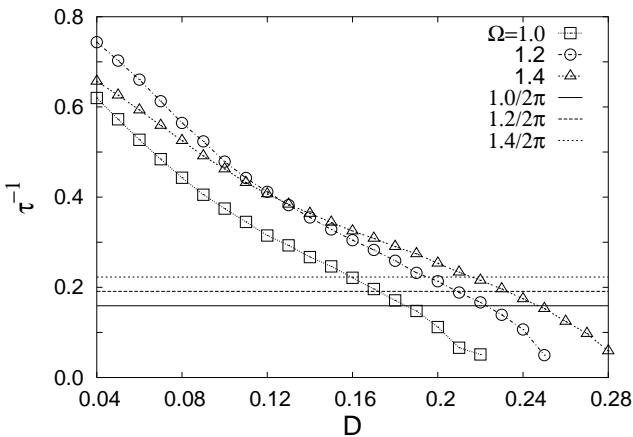}}}
\caption{Inverse relaxation time $\tau^{-1}$ vs the 
disorder strength $D$. 
The crossing points with the lines $1.0/2\pi, 1.2/2\pi$, and $1.4/2\pi$ yield 
the values of $D_m$ at the corresponding driving frequency.}
\label{fig:tau}
\end{figure}
Figure~\ref{fig:tau} shows the inverse relaxation time $\tau^{-1}$
versus the disorder strength 
$D$ for $I_0=0.8$ and $\Omega=1.0,~1.2$, and $1.4$.  
The intrinsic time scale $\tau$ of the system 
is observed to increase indefinitely as the disorder strength $D$ approaches 
the critical value $D_c$ beyond which synchronization disappears. 
The horizontal lines in Fig.~\ref{fig:tau} describe the time-scale matching 
conditions at various driving frequencies: the value of $D$ at the 
crossing point corresponds to $D_{m}$ 
at which the enhancement of the response is maximized.
For example, for $\Omega=1.0,~1.2$, and $1.4$, the crossing points lead to 
$D_m \approx 0.19,~0.21$, and $0.22$, respectively, which are consistent with 
the values obtained both analytically and numerically  [see Figs.~\ref{fig:f1} 
and \ref{fig:histo_reso}]. 
We thus conclude that quenched disorder may also enhance the 
coherent response of the system via the mechanism of the time-scale matching.

In summary, we have explored the effects of quenched disorder on the 
coherent response in the driven system of coupled oscillators. 
We have investigated the interplay between quenched disorder and external 
periodic driving, with particular attention to the possibility of 
resonance behavior.  The phase velocity is probed as the response of the 
system; revealed is the enhancement of the fraction of the oscillators
locked to the external driving, exhibiting resonance behavior.  
This provides the first observation of the resonance behavior induced by 
quenched disorder.  
In a biological system such as synchronous fireflies, different firing 
frequencies of fireflies may be regarded as the quenched disorder.  
Our results are applicable to expect those different 
firing frequencies (instead of same ones) may enhance the coherent response 
of the system.  

We thank M.Y. Choi, H. Park, and B.J. Kim for useful discussions.


\begin{thebibliography}{20}
\bibitem{ref:SR1}
R. Benzi, A. Sutera, and A. Vulpiani, J. Phys. A {\bf 14}, L453 (1981); 
R. Benzi, A. Sutera, G. Parisi, and A. Vulpiani, SIAM J. Appl. Math. {\bf 43},
565 (1983).

\bibitem{ref:SR2}
B. McNamar and K. Wisenfeld, Phys. Rev. A {\bf 39}, 
4854 (1989); L. Gammaitoni, F. Marchesoni, E. Menichella-Saetta, and 
S. Santucci, Phys. Rev. Lett. {\bf 62}, 349 (1989).

\bibitem{ref:SR3}
L. Gammaitoni, P. H{\"a}nggi, P. Jung, and F. Marchesoni, Rev. Mod. Phys. 
{\bf 70}, 223 (1998). 

\bibitem{ref:Schmitt}
S. Fauve and F. Heslot, Phys. Lett. {\bf 97 A}, 5 (1983).

\bibitem{ref:ringlaser}
B. McNamara, K. Wiesenfeld, and R. Roy, Phys. Rev. Lett. {\bf 60}, 
2626 (1988).

\bibitem{ref:crayfish}
K. Wiesenfeld, D. Pierson, E. Pantazelou, C. Dames, and F. Moss, Phys. Rev. Lett. {\bf 72}, 
2125 (1994).

\bibitem{ref:Kuramoto}
Y. Kuramoto, in {\it Proceedings of the International Symposium on 
Mathematical Problems in Theoretical Physics}, edited by H. 
Araki (Springer-Verlag, New York, 1975); Prog. Theor. Phys. Suppl. 
{\bf 79}, 223 (1984); 
Y. Kuramoto and I. Nishikawa, J. Stat. Phys. {\bf 49}, 569 (1987). 


\bibitem{ref:Kuramoto_Sakaguchi}
H. Sakaguchi and Y. Kuramoto, Prog. Theor. Phys. {\bf 76}, 576 (1986).

\bibitem{ref:Wiesenfeld_Colet_Strogatz}
K. Wisenfeld, P. Colet, and S.H. Strogatz, Phys. Rev. Lett. {\bf 76}, 
404 (1996); Phys. Rev. E {\bf 57}, 1563 (1998).

\bibitem{ref:period_synch} M.Y. Choi, Y.W. Kim, and D.C. Hong, Phys. Rev. E {\bf 49}, 3825 (1994). 

\bibitem{ref:Shapiro} S. Shapiro, Phys. Rev. Lett. {\bf 11}, 80 (1963); 
S. Shapiro, A.R. Janus, and S. Holly, Rev. Mod. Phys. {\bf 36}, 223 (1964). 


\bibitem{ref:Heun} See, e.g., R.L. Burden and J.D. Faires, 
{\it Numerical Analysis} (Brooks/Cole, Pacific Grove, 1997), p. 280.

\end{thebibliography}
\end{document}